# Hot ion generation from nanostructured surfaces under intense, femtosecond irradiation

S. Bagchi, P. Prem Kiran, M.K. Bhuyan, S. Bose, P. Ayyub, M. Krishnamurthy and G. Ravindra Kumar*

Tata Institute of Fundamental Research, 1 Homi Bhabha Road, Colaba, Mumbai-400005, India

Abstract

We present the effect of a nanostructured surface on the emission of ions and electrons from intense (5-36 Petwatt per sq.cm) femtosecond laser produced plasmas. Electrons from optically polished copper targets coated with copper nanoparticles (CuNP) are observed to be hotter than those from uncoated polished targets. A nearly two-fold enhancement is observed for ions in the range 14-74 keV, while ion yield decreases by a factor of 2 in the 74-2000 keV range. The total ion yields measured using a large area Faraday cup are more from CuNP targets than those from polished Cu targets, indicating increased ion beam divergence due to surface modulations.

.

* Author for correspondence:
Tata Institute of Fundamental Research,
1 Homi Bhabha Road, Colaba, Mumbai-400005, India
Fax : +91-22-22804610/11 ;
e-mail: grk@tifr.res.in

1. **Introduction**

The advent of ultrashort, intense solid state lasers has enabled the production of high quality particle beams with high energy and brightness[1-2]. Not only is the physics of these ultrashort duration beams very interesting, but their potential applications in areas like inertial confinement fusion, accelerators, isotope production for medicine and surface modifications are extremely important for future technologies[3]. When an intense laser pulse interacts with the target, rapid ionization occurs at the beginning of the laser pulse and the electrons in the plasma then absorb the light by a variety of mechanisms like resonance absorption (RA)[4] and vacuum heating (VH)[5]. The electrons which gain energy through these mechanisms form a distinct 'hot' bunch well separated in energy from the colder electrons in the bulk plasma. The hot electrons radiate part of their energy via x-ray emission and also transfer their energy to the ions in the plasma. Ion acceleration occurs when the hot plasma expands and an electric field is created due to charge separation. Ion acceleration in plasma with multiple electron temperatures has been explored in simulations[6,7] and in different experimental situations. The multiplicity of species and how it leads to deviations from self-similarity in the expansion has been examined[6]. Many studies have reported the systematics of particle production from solid[2,3,8], cluster[9] and microdroplet[10] plasmas. Efficient coupling of laser energy into the short-lived plasma thus plays a crucial role in ion generation. Enhanced coupling of laser energy to the plasma has been achieved by the creation of a preplasma before the arrival of the main pulse, by modifying target composition in the case of clusters[9] and by the introduction of sub laser wavelength surface modulations on solid targets.[11,12] In fact, the introduction of a structured surfaces has been shown to increase the coupling of laser energy by as much as 80 % as compared to that for polished surfaces. A general question that may be posed is whether ion energies can similarly be increased by optimization of the target properties; A more pertinent question is whether the same optimization scheme will work for enhanced generation of hot electrons as well as higher energy ions. This is based on the expectation that hotter electrons (enhanced coupling of laser energy) should lead to hotter ions. Is this borne out in experiments? In this letter we report systematic study of the influence of surface modulations on characteristics of the emitted ions from a femtosecond laser produced plasma. With this perspective, here we present results for nanoparticle coated targets.

2.  **Experimental Details**

The experiment (Fig. 1) is performed with focused 50 fs, p-polarized laser pulses (1-6 mJ) from a 806 nm, 10Hz, Ti-Saphhire laser (THALES LASER, ALPHA 10). The target is a polished copper block (50 mm × 50 mm × 5 mm), half of which is coated with a thick layer of copper nanoparticles (CuNP) with an average size of 15 nm. The nanoparticles are deposited using high-pressure DC magnetron sputtering technique.[13] The crystallite size is determined from the Scherrer broadening of the Cu [111] x-ray diffraction line. The partially coated target ensures exactly the same laser and detector conditions for measurements from both surfaces. The typical coating thickness of the NPs is about 0.2 μm which is large compared to the optical skin depth of a few nanometers. The base pressure of the chamber is $10^{-6}$ torr. The target is scanned across the laser beam to ensure that every laser pulse hits a fresh region. The laser is focused by a gold coated off axis parabolic mirror (OAP), in an f/4 focusing geometry with a spot size (FWHM) of 10 μm giving peak intensites in the $0.5 - 3.6 \times 10^{16}$ Wcm$^{-2}$ range. The hard x-ray spectra in the 20 – 200 keV range are measured for a total of 3000 – 4000 laser shots using a calibrated NaI(Tl) scintillation detector shielded by 1.5 cm thick lead bricks, coupled with a multichannel analyzer. The detector is gated in time with the laser pulse and the signal is collected only in a time window of 30 μs to ensure background-free acquisition. Bremsstrahlung temperature fits are done using the data above 50 keV, where the transmission is close to 100%. Count rates are kept at less than 0.1 per laser shot (to prevent pile up) by introducing suitable lead apertures and restricting the solid angle to 50 – 80 μsr. The ions emitted normal to the target surface are detected by a channel electron multiplier (CEM) and a Faraday cup (FC) arrangement. The CEM, kept at a distance of 97 cm from the focal spot directly views the plasma plume subtending a solid angle of 26 msr. The FC (230 cm$^2$ area) is used to collect all the emitted charge particles from the plasma to obtain the total ion flux. The FC, made of 11μm thick Aluminium foil is biased at +100 V to arrest secondary electrons. It stops all the heavier ions, while transmitting higher energy electrons.[14]

3. **Results and inferences**

Fig. 2 shows the ratio of total x-ray yield from CuNP coated surface to that from the polished Cu surface at different input laser energy corresponding to the intensities in the range of 0.9 – 3.6 × $10^{16}$ Wcm$^{-2}$. The yield from NPs is approximately 4 times that from the polished Cu surface, up to an intensity of 2.5 × $10^{16}$ Wcm$^{-2}$ after which it gradually decreases to approximately 1.5 times. The bremsstrahlung (hard x-ray) spectra from both the polished Cu and CuNP coated surface at the input intensity of 3 × $10^{16}$ Wcm$^{-2}$ are shown as insets (a) and (b) respectively. We observe a clear enhancement in the hard x-ray spectrum of CuNP, indicating hotter electrons in the nanoplasma, in tune with previous results.[11] The total (integrated) energy of x rays emitted in the range of 30 – 200 KeV for polished Cu and NP surface is 2.9×$10^4$ KeV (2.9× $10^{-9}$ mJ) and 4.4×$10^4$ KeV (7.05× $10^{-9}$ mJ) respectively. The hot electron distribution in Cu plasma is a single Maxwellian with a temperature of 9.3 ± 1.1 KeV, whereas the CuNP plasma shows two-temperatures: 9.5 ± 1.7 KeV and 33.9 ± 6.4 KeV, indicating enhanced coupling of laser energy into NP coated surface when compared to the polished surface. For p-polarized light, the hot electrons are generated mainly by resonance absorption (RA) under our conditions. The hot electron temperature, $T_{hot}$, can be estimated following the scaling law:[16] $T_{hot} = 14 T_c^{0.33} (I\lambda^2)^{0.33}$, where $T_c$ is the background electron temperature in keV, $I$ is the intensity of the laser in units of $10^{16}$ Wcm$^{-2}$ and $\lambda$ is the wavelength in microns. According to this scaling law, for a $T_c$ of 0.15 keV (estimated for our intensities), we get a $T_{hot}$ of 9.2 keV. This temperature is close to that observed for the polished target and the lower of the two components observed in the nanoparticles coated targets. This also agrees with earlier measurements.[11,17] The hotter component of 33 keV agrees with our earlier studies and is attributed to local field enhancements of the laser light or alternately surface plasmon excitation facilitated by the nanoparticle coating.[11,12]

Time of flight spectra (inset of Fig. 3a) of the ions clearly show differences in energies observed from the NP-coated surface and polished Cu. The polished Cu surface, at lower input intensities, shows two main features representing two different ion species of 50 – 80 and 220 – 257 KeV energy. With increasing input energy, the two peaks merge, giving a single distribution of the ions in the 220 – 340 KeV energy range. In contrast, the CuNP surface gives three distinct peaks at 25, 46, and 324 KeV at low incident laser energy. This evolves into a two-peak feature indicating the presence of two sets of ions (B and C in inset of Fig. 3a) of energies 46 – 61 and 238 – 324 KeV. From the n(E) – E spectrum, where n(E) is the number of ions emitted (Fig. 3b) within the energy range of E to E+dE, a clear presence of two distributions of ions is visible. Though the number of particles emitted from the both the Cu and CuNP targets is same in the low energy range 4 – 16 KeV (region A of Fig. 3a), the number of particles emitted from the nanoplasma is nearly 1.5 times more in the intermediate energy (16 – 75 KeV) range (region B of Fig. 3a). Above 75 KeV (region C of Fig. 3a) the number is lower by a factor of 2. At all the incident energies used in our studies, the maximum ion energy as well as the average energy from the CuNP is observed to be lower than the corresponding values from polished surface. The ratio of the ion yield, from nanostructured surface to that from polished Cu ($Y_{CuNP}/Y_{Cu}$) is in the range of 0.84 – 0.89 indicating lower ion yield from the Cu nanoplasma. The cutoff ion energy from both the CuNP and Cu surface over the input laser energy is shown in Fig 3b. The cutoff ion energy from the Cu plasma increases from 1.58 to 3.80 MeV as the laser intensity increases from 0.9 to $3.6 \times 10^{16}$ Wcm$^{-2}$. The corresponding values are 1.0 to 2.0 MeV for CuNP plasma indicating a reduction of around 50 % in the high energy ion emission from the CuNP plasma. Above an intensity of $3.2 \times 10^{16}$ Wcm$^{-2}$, the cutoff ion energy from the CuNP plasma is almost the same as that for Cu, indicating the damage to the NP coating, in agreement with our previous results.[15] We also examined the possibility that protons could be preferentially enhanced in their energy in the nanoparticle coated target, but that is ruled out by the measured time of flight spectra, the details of which wil be reported in a longer paper.

This clearly shows that though more laser energy is coupled to the NP-coated surface than the polished surface, the ion emission from the NP coated surface gets preferentially enhanced in an intermediate regime. The highest ion energy decreases for the NP coated surface. This is in contradiction to the usual expectation that the generation of hotter electrons should lead to hotter ion emission.[7,18] The CEM measurements show that the total particle yield as well as the maximum ion energy is always lower than the polished target over the entire laser energy range used. To get a measure of the overall ion yield, we measured the total ion currents from the two types of targets at a location closer to the target, using a large area Faraday cup (FC) subtending a solid angle of 2.36 Sr to the plasma focus. The total charge accumulated by the FC (Fig. 4b) shows that the total particle yield is 25% more in case of CuNP targets compared to polished Cu targets. This measuremen is contrary to the result of the low solid angle CEM measurement, and these two results indicate that the ion beam from the nanoparticle coated targets has a greater divergence.

To look at our results from a comparative perspective, let us first consider the ion emission from the polished (unstructured) copper targets. The hot electrons produced by the intense laser, together with the colder electrons form a two–temperature plasma[6]. The 'self-similar' expansion of one electron temperature plasma gives rise to ions that eventually emerge with the speed of sound in the plasma medium[19]. The two-temperature plasma on the other hand gives rise to a sheath formation at a point in the density profile across which the ions are accelerated and these emerge as the hot ions.[6] The field developed across the sheath can be expressed as $E_{accl} = k_B T_e /e[\max (L_n, \lambda_D)]$, where $L_n$ is the local scale length of the expanding plasma and $\lambda_D$ is the Debye length. [2,20] The magnitude of the accelerating electric field $E_{accl}$ depends on the temperature of the escaping hot electrons as well as on the sheath thickness, which in turn depends on the plasma electron density and temperature. The model assumes isothermal expansion which may not be strictly valid, but the general features are well reproduced even under this approximation.

Following Mora[17], the maximum energy of the ions can be estimated by $E_{max} = 2E_0 [\ln(2\tau)]$, where $E_0 = Z k_B T_h$, $\tau = 0.43\, \omega_{pi}$ and the ion plasma frequency $\omega_{pi} = (n_{e0} Z e^2 / m_i \epsilon_0)^{1/2}$, $n_{e0}$ being the unperturbed electron density and Z is the charge state of the ions. Given our observed cut off ion energy of 2.15 MeV, and assuming an average ionization state Z = 4, $n_{e0} = n_{cr}$ and a hot electron temperature $T_h = 9.3$ keV, we get $\tau = 0.29$ indicating that the long time approximation is valid (much longer than the laser pulse duration). The maximum velocity comes out to be $2.51 \times 10^6$ m/sec. This also matches well with the observed velocities of $2.44 \times 10^6$ m/sec in the experiment. While applying these derived parameters for CuNP we get maximum ion energies of 2.15 MeV and 7.7 MeV for the two hot electron temperature as observed in bremsstrahlung spectra in contradiction to the observed maximum ion energy of 2.0 MeV (Fig. 3b) clearly indicating that the accelerating potential for ion expansion is suppressed in case of nanoparticles. So the key question that we need to address is the suppression of ion energies in comparison to the observed energies arising in polished targets.

We are investigating the reduction in the maximum ion energy as well as the multispecies ion acceleration using PIC simulations.[21] In this letter, we offer some plausible physical explanations. Firstly, the nanoparticle coating causes significant deviation from the assumption of planar 1D expansion, clearly evident in our Faraday cup measurements of the divergence (the nanoparticles are 'smearing the ions). It is well known that in multidimensional (ex: spherical) expansion the maximum energy is lower than the 1 D case.[22] The nanoparticles act as the first sources of the plasma because of the enhanced laser intensities in the immediate vicinity of each particle.[11] If we take the critical plasma density ($10^{21}$ cm$^{-3}$) and $T_h$ =9.3 keV, the Debye length is 17 nm, which is of the same size as the nanoparticles. The actual Debye length for colder temperatures achieved on the rising edge of the laser pulse is much smaller. The plasma is likely to expand away from these hot spots with significant nonplanarity, guided mainly by the initial shape of these ellipsoidal particles. The effective Debye length is thus smaller and leads to lower ion velocities as given by eq. It is also possible that the nanoparticles perturb the plasma by 'lightning rod' effects[10] and these can prevent charge build up and disturb sheath formation which results in lower sheath voltage and hence lower expansion. It is known from the numerical simulations as well as from the experimental observations that the particle emission happens mostly normal to the surface and the nonplanar plasma expansion thus gives rise to the increase beam divergence. The reduction in the maximum energy of the protons and ions from the laser produced plasmas leading to the increased scale length resulting in divergent ion and proton beams due to the surface modulations have been observed.[23,24] All these arguments are supported by our recent observations of the survival[15] of these nanoparticles at intensities as high as $2 \times 10^{16}$ Wcm$^{-2}$.

The present study is significant because it shows that hot electron generation ***need not*** cause hotter ion emission. We point out that, in general, optimization of each signal from the plasma requires an understanding of the actual dynamics of the process. In our case, hot electron generation is enhanced by the local field enhancements which increase the effective light intensity. This process takes place on the time scale of the laser pulse (tens of fs). The ion acceleration, however, depends on a subsequent process namely the plasma expansion and energy exchange. We believe that this is significantly modified by the same nanoparticles in a counteracting manner and leads to enhanced number of ions in the intermediate range (14 – 74 KeV) and reduces the high energy ions above 74 KeV.

In summary, we have presented ion energy and flux measurements from plane polished targets as well as those coated by copper nanoparticles. A significant and surprising observation is that the nanoparticles coated targets give hotter electrons but colder ions. The enhancement of x-ray emission and selective suppression of high energy ion emission can prove very useful in designing brighter hard x-ray sources with reduced damage from ion debris[24]. We expect that this result will provoke a discussion of the physics involved and help in the design of femtosecond laser driven x-ray and ion sources.

**Figure Captions:**

**Figure 1**: Schematic of the experimental arrangement.

**Figure 2**: Ratio of total x-ray from CuNP and Cu surface with laser energy. Inset shows the Bremsstrahlung spectra of CuNP coated and polished Cu surfaces at $3 \times 10^{16}$ Wcm$^{-2}$.

**Figure 3:** The (a) n(t) – t and n(E) – E spectra of ions emitted from CuNP and Cu surfaces at laser energy of 5 mJ (~ $3 \times 10^{16}$ Wcm$^{-2}$). Inset shows the time of flight spectrum. Vertical lines indicate three different regions of ion energies. (b) Ion Cutoff energies from Cu and CuNP as a function of input laser Energy.

**Figure 4:** Ratio of the total charge collected by Faraday Cup from CuNP and Cu.

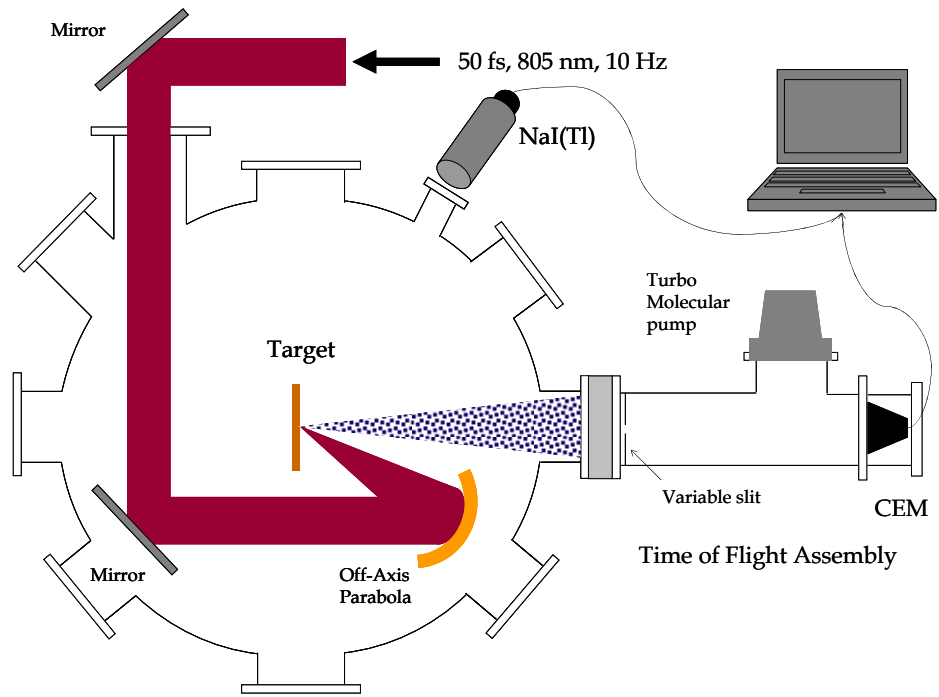

**Figure 1**: Schematic of the experimental arrangement.

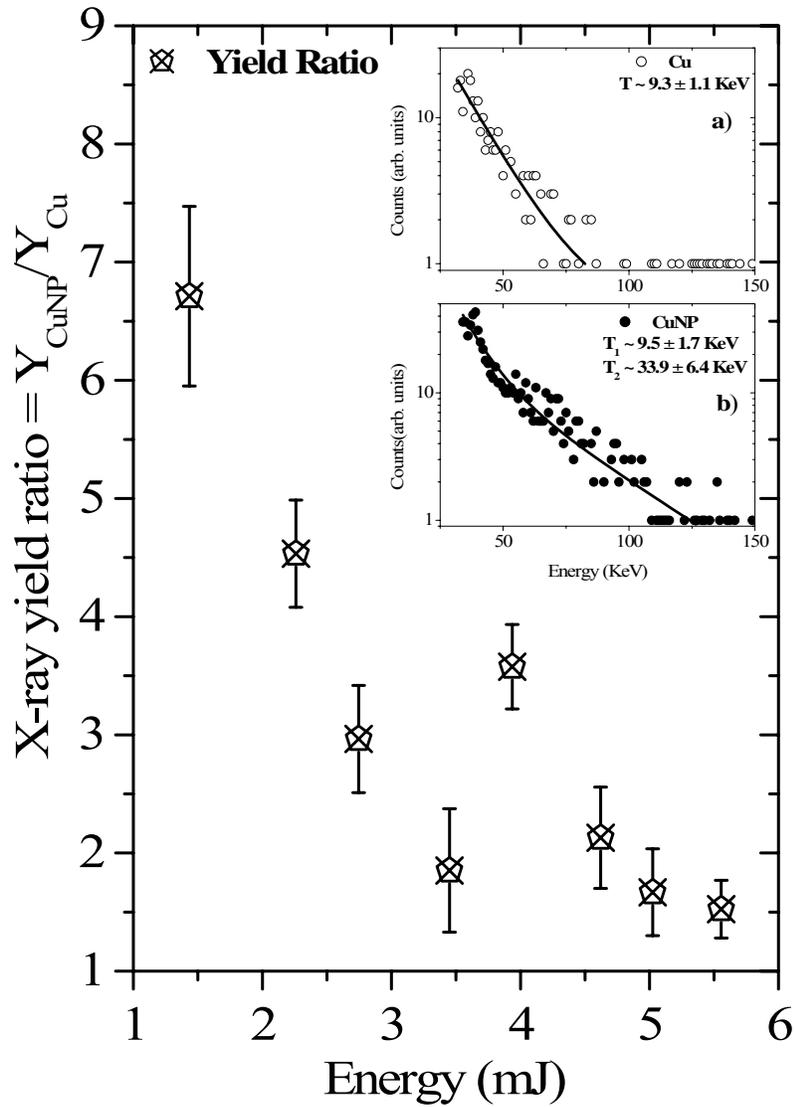

**Figure 2:** Variation of the ratio of total x-ray yields from CuNP ($Y_{CuNP}$) and Cu surface ($Y_{Cu}$) with laser energy. Inset shows Bremsstrahlung spectra of CuNP coated and polished Cu surfaces at $3 \times 10^{16}$ Wcm$^{-2}$.

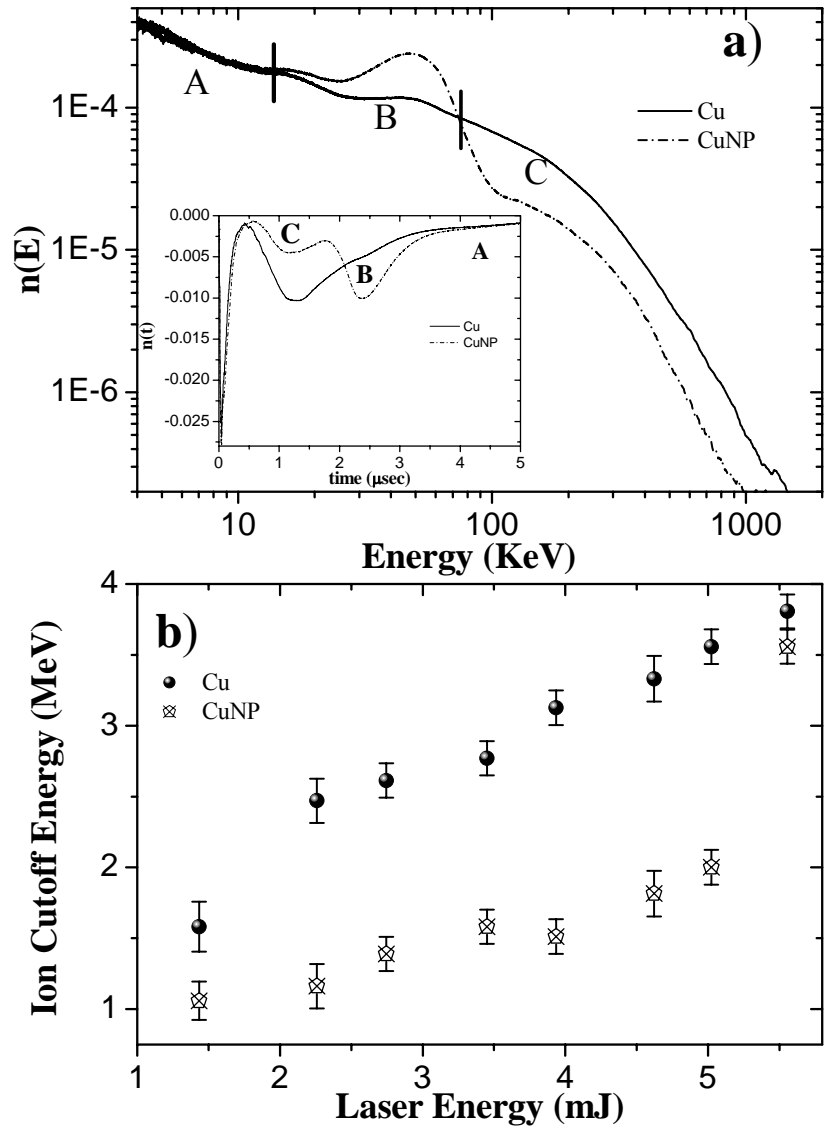

**Figure 3**: The (a) n(t) – t and n(E) – E spectra of ions emitted from CuNP and Cu surfaces at laser energy of 4.6 mJ (~ $2.7 \times 10^{16}$ Wcm$^{-2}$). Inset shows the time of flight spectrum. Vertical lines indicate three different regions of ion energies. (b) Ion Cutoff energies from Cu and CuNP as a function of input laser Energy.

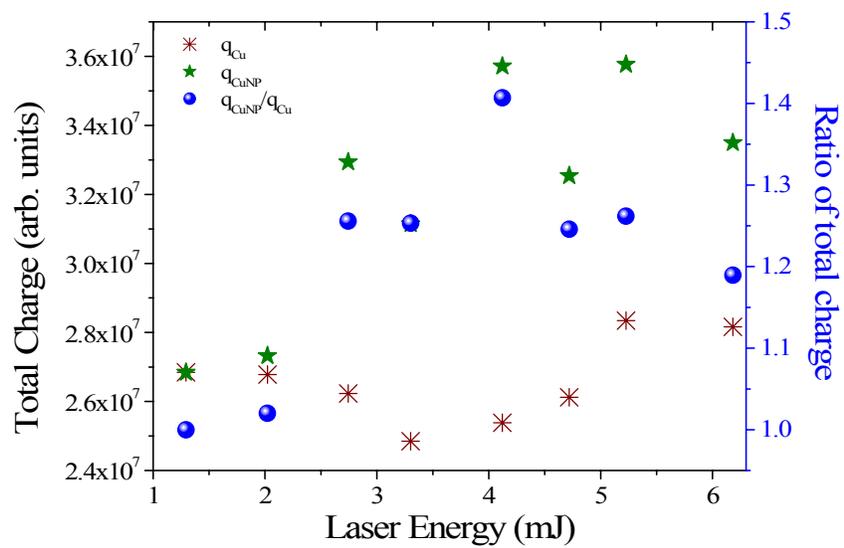

**Figure 4:** Ratio of the total charge collected by Faraday Cup from CuNP and Cu.